\documentstyle[aps,preprint,amssymb,psfig]{revtex}
\newcommand{\cbeta}{{\cal B}}
\newcommand{\ib}{{\rm int}}
\newcommand{\lb}{{\rm lim}}
\begin{document}
\draft   
\title{Classification of phase transitions in small systems}
\author{Peter Borrmann, Oliver M{\"u}lken and Jens Harting}
\address{Department of Physics, Carl von Ossietzky University
Oldenburg, D-26111 Oldenburg, Germany}
\date{\today}
\maketitle
\begin{abstract}
We present a classification scheme for phase transitions in finite
systems like atomic and molecular clusters based on the Lee-Yang zeros
in the complex temperature plane. In the limit of infinite particle
numbers the scheme reduces to the Ehrenfest definition of phase
transitions and gives the right critical indices. We
apply this classification scheme to  
Bose-Einstein condensates in a harmonic trap as an example of 
a higher order phase transitions in a finite system and to small 
Ar clusters.
\end{abstract}
\pacs{PACS numbers: 05.70.Fh,64.60.Cn,64.70.-p}
{\sl Small systems do not exhibit phase transitions.} Following
Ehrenfest's definition this statement is true for almost all small
systems. Instead of exhibiting a sharp peak or a discontinuity in the
specific heat at some well defined critical temperature the specific
heat shows a more or less smooth hump extending over some finite
temperature range. For example for the melting of atomic clusters this
is commonly interpreted as a temperature region where solid and liquid
clusters coexist \cite{Berry98a,Schmidt98} and as a finite-system
analogue of a first order phase transition. Proykova and Berry
\cite{proykova} interpret a structural transition in TeF$_6$ clusters as
a second order phase transition. A common way to investigate if a
transition in a finite system is a precursor of a phase transition in
the corresponding infinite system is to study the particle number
dependence of the appropriate thermodynamic potential \cite{mour}.
However, this approach will fail for all system types where the nature
of the phase transition changes with increasing particle number which
seems to be the case e.g. for sodium clusters \cite{sodium} or
ferrofluid clusters \cite{jcptls}.  For this reason a definition of
phase transitions for systems with a fixed and finite particle number
seems to be desirable. The only recommended feature is that this
definition should reduce to the Ehrenfest definition when applied to
infinite systems for systems where such limits exist. An approach in 
this direction has been made by studying the topological structure 
of the n-body phase space and {\sl a hypothetical  definition} based  
on the inspection of the shape of the caloric curve \cite{Gross}. A
mathematical more rigid investigation giving the sufficient and 
necessary conditions for the existence of van der Waals-type loops 
has been given by Wales and Doye \cite{Doye}.

Our ansatz presented in this letter is based on earlier works of Lee and
Yang \cite{Yang1952a} and Grossmann {\sl et al.} \cite{Gross1967} who
gave a description of phase transitions by analyzing the distributions
of zeros (DOZ's) of the grandcanonical $\Xi(\beta)$ and the canonical
partition function $Z(\beta)$ in the complex temperature plane.  For
macroscopic systems this analysis merely contributes a sophisticated
view of the thermodynamic behavior of the investigated system.  We will
show that for small systems the DOZ's are able to reveal the
thermodynamic secrets of small systems in a distinct manner. In the
following we restrict our discussion to the canonical ensemble and
denote complex temperatures by $ \cbeta = \beta + i \tau$ where $\beta$
is as usual $1/k_{\rm B} T$\cite{extension}. 

In the case of finite systems one must not deal with special
considerations regarding the thermodynamic limit. We write the canonical
partition function $Z(\cbeta)~=~\int {\rm d} E \Omega(E)\exp(-\cbeta
E)$, with the density of states $\Omega(E)$, as a product
\begin{equation}
Z(\cbeta) = Z_\lb(\cbeta) Z_\ib (\cbeta)
\end{equation}
where $Z_\lb(\cbeta)$ describes the limiting behavior of $Z(\cbeta)$ for
$T \rightarrow \infty $ imposing $\lim_{\cbeta \to 0}
Z_\ib(\cbeta)~=~1$.  In general, $Z_\lb(\cbeta)$ will not depend on the
interaction between the particles or the particle statistics but it will
depend on the external potential imposed. E.g. for $N$ particles in a
$d$-dimensional harmonic trap we have $Z_\lb(\cbeta) \approx \cbeta^{-d
N}$ and for a $d$-dimensional gas $Z_\lb(\cbeta) \approx \cbeta^{-d
N/2}$. In the following we will assume that $Z_\lb(\cbeta)$ has no zeros
except at $\cbeta=\infty$. Then the zeros of $Z(\cbeta)$ are the same as
those of $Z_\ib(\cbeta)$.  Applying the product theorem of Weierstrass
\cite{Titchmarsh} the canonical partition function can be written as a
function of the zeros of $Z_{\ib}(\cbeta)$ in the complex temperature
plane. Because $Z(\cbeta)$ is an integral function its zeros $\cbeta_k =
\cbeta_{-k}^{*} = \beta_k + i \tau_k\;\; (k \in \mathbb{N})$  are
complex conjugated 
\begin{eqnarray} \label{zerospart}
Z(\cbeta)&=&Z_\lb(\beta) Z_\ib(0) \exp \left(
\cbeta \partial_{\cbeta} \ln Z_\ib(0) \right) \\
&&\times \prod_{k \in \mathbb{N}} 
\left(1 - \frac{\cbeta}{\cbeta_k}\right)
\left(1 - \frac{\cbeta}{\cbeta_k^{*}}\right)
\exp\left(\frac{\cbeta}{\cbeta_k}+ \frac{\cbeta}{\cbeta_k^{*}}
\right)  \nonumber \\
&=& Z_\lb(\beta)
\prod_{k \in \mathbb{N}} \left( \frac{\tau_k^2 + (\beta_k -
\cbeta)^2}{\beta_k^2
+ \tau_k^2} \right)  \exp\left( \frac{2 \cbeta \beta_k}{\beta_k^2
+ \tau_k^2}\right) \;\; . \nonumber
\end{eqnarray}
The free energy $F(\cbeta)=-\frac{1}{{\cal B}} \ln(Z({\cal B}))$ is
analytic, i.e. it has a derivative at every point,
everywhere in the complex temperature plane except at the zeros of
$Z(\cbeta)$.  If the zeros are dense on lines in the complex plane,
different phases are represented by different regions of holomorphy of
$F(\cbeta)$ and are separated by these lines in the complex temperature
plane. The DOZ contains the complete thermodynamic information about the
system and all desired thermodynamic functions are derivable from it.
The calculation of the specific heat $C_V(\cbeta)$ by standard
differentiation yields 
\begin{equation} \label{heat}
C_V(\cbeta) = C_{{\rm lim}}(\cbeta) - k_{\rm B} \cbeta^2  
\sum_{k \in \mathbb{N}}
\left[
  \frac{1}{ (\cbeta_k-\cbeta)^2}
+ \frac{1}{ (\cbeta_k^{*}-\cbeta)^2}
\right] \;\; .
\end{equation}
Zeros of $Z(\cbeta)$ are poles of $F(\cbeta)$ and $C_V(\cbeta)$.
As can be seen from eq.~(\ref{heat}) the major contributions 
to the specific heat come from zeros close to the real axis, and
a zero approaching the real axis infinitely close causes 
a divergence in the specific heat.
 
\begin{figure}
\centerline{\psfig{file=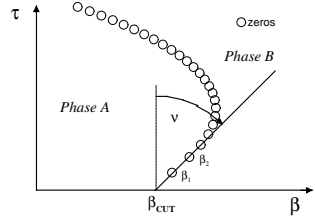,height=3cm,angle=270}}
\caption{Schematic plot of the DOZ illustrating the 
definition of the classification parameters given in the text.}
\label{figure1}
\end{figure}

In the following we will give a discretized version of the
classification scheme of Grossmann {\sl et al.}\cite{Gross1967}.
To characterize the DOZ close to the real axis let us assume
that the
zeros lie approximately on a straight line.  The crossing angle of this
line with the imaginary axis (see Fig.~\ref{figure1}) is then $\nu =
\tan\gamma$ with $\gamma = (\beta_2 - \beta_1)/(\tau_2-\tau_1)$. The
crossing point of this line with the real axis is given by 
$\beta_{\rm cut} = \beta_1 - \gamma \tau_1$. 
We define the discrete line density $\phi$ as a
function of $\tau_k$ as the average of the inverse distances between
$\cbeta_k$ and its neighboring zeros 
\begin{equation}
   \phi(\tau_k) =  \frac{1}{2} \left(  
        \frac{1}{|{\cal B}_{k} - {\cal B}_{k-1}|} 
      + \frac{1}{|{\cal B}_{k+1} - {\cal B}_{k}|} 
        \right) \;\;(k =2,3,4,\ldots)\;\;.
\end{equation}
Guide lined by the fact that the importance of the contribution of a
zero to the specific heat decreases with increasing $\tau$ we
approximate $\phi(\tau)$ in the region of small $\tau$ by a simple power
law $\phi(\tau) \sim \tau^{\alpha}$. A rough estimate of $\alpha$
considering only the first two zeros yields 
\begin{equation}
\alpha = \frac{\ln\phi(\tau_3) - \ln\phi(\tau_2)}
          {\ln\tau_3 - \ln\tau_2}\;.\;
\end{equation}
Together with $\tau_1$, the imaginary part of the zero closest to the
real axis, the parameters $\gamma$ and $\alpha$ classify the DOZ. As
will be shown below, the parameter $\tau_1$ is the essential parameter
to classify phase transitions in small systems.  For a {\sl true} phase
transition in the Ehrenfest sense we have $\tau_1 \rightarrow 0$. For
this case it has been shown \cite{Gross1967} that a phase transition is
completely classified by $\alpha$ and $\gamma$.  In the case $\alpha =
0$ and $\gamma = 0$ the specific heat $C_V(\beta)$ exhibits a
$\delta$-peak corresponding to a phase transition of first order. For $0
< \alpha < 1$ and $\gamma =0$ (or $\gamma \neq 0$) the transition is of
second order. A higher order transition occurs for  $1 < \alpha $ and
arbitrary $\gamma$.  This implies that the classification of phase
transitions in finite systems by $\gamma$, $\alpha$, and $\tau_1$, which
reflects the finite size effects, is a straightforward extension of the
Ehrenfest scheme.

The imaginary parts  $\tau_i$ of the zeros have  a simple
straightforward interpretation in the quantum mechanical case. By going
from real temperatures $\beta=1/(k_{\rm B} T)$ to complex temperatures
$\beta + i \tau/\hbar$ the quantum mechanical partition function can be
written as
\begin{eqnarray}
Z(\beta+i \tau/\hbar)&=&{\rm Tr}\left( 
\exp(-i \tau {\rm H}/\hbar)
\exp(-\beta {\rm H}) \right) \label{canstate}\\
&=& \langle \Psi_{{\rm can}} \mid 
\exp\left(-i \tau {\rm H}/\hbar \right)
\mid \Psi_{{\rm can}} \rangle \\ 
&=& \langle \Psi_{{\rm can}}(t=0)\mid \Psi_{{\rm can}}(t=\tau) 
\rangle \;\;, \nonumber
\end{eqnarray}
introducing a ''{\sl canonical state}'' $\mid \Psi_{{\rm can}} \rangle = 
\sum_i \exp(-\beta\epsilon_i/2) \mid \phi_i \rangle \quad$ ,
which is  the sum of all eigenstates of the system appropriately
weighted by the Boltzmann factor.  Within this picture a zero of the
partition function occurs at times $\tau_i$ where the overlap of a time
evoluted canonical state and the initial state vanishes. This resembles
a correlation time, but some care is in order here. The time $\tau_i$ is
not connected to a single system, but to an ensemble of infinitely many
identical systems in a heat bath, with a Boltzmann distribution of
initial states.  Thus, the times $\tau_i$ are those times after which
the whole ensemble loses its memory. 

Equation~(\ref{canstate}) is nothing but the canonical ensemble average
of the time evolution operator $\exp\left(-i \tau {\rm H}/\hbar
\right)$. Following Boltzmann the ensemble average equals the long time
average which was proven quantum mechanically by
Tasaki~\cite{tasaki:1998}.  Therefore $\tau_i$ indeed resembles times
for which the long time average of the time evolution operator vanishes. 

The observation of Bose-Einstein condensation in dilute gases of finite
number ($ \sim 10^3-10^7$) of alkali atoms in harmonic traps
\cite{Anderson1995a} has renewed the interest in this phenomenon which
has already been predicted by Einstein \cite{Bose1924a} in 1924.  The
number of condensed atoms in these traps is far away from the
thermodynamic limit, raising the  interesting question how the order of
the phase transition changes with an increasing number of atoms in the
condensate. For this reason we treat the Bose-Einstein condensation in a
3-dimensional isotropic harmonic trap ($\hbar=\omega=k_{\rm B}=m=1$) as
an example for the application of the classification scheme given above.

For non-interacting bosons the occupation numbers
of an eigenstate $|i\rangle$  and $N+1$ particles
can be evaluated by a simple recursion \cite{bose1}
\begin{equation} \label{recocc}
\eta_i(N+1,\cbeta) = \frac{Z_N(\cbeta)}{Z_{N+1}(\cbeta)} 
\exp(-\cbeta \epsilon_i) (\eta_i(N,\cbeta)+1) \; .
\end{equation}
Since the particle number is a conserved quantity in the canonical
ensemble the direct calculation of the normalization factor 
can be omitted by using the relation
\begin{equation} \label{norm}
\frac{Z_N(\cbeta)}{Z_{N+1}(\cbeta)} = 
\frac{N+1}{\sum_{i=0}^{\infty} \exp(-\cbeta \epsilon_i)
(\eta_i(N,\cbeta)+1)} \;.
\end{equation}
Since $Z_N(\cbeta)$ is an exponentially decreasing function in
$\beta$ it is a difficult numerical task to calculate its zeros 
directly. 
Zeros of the partition function are reflected by poles of the ground
state occupation number
\begin{equation}
\eta_0(N,\cbeta)=-\frac{1}{\cbeta}
\frac{\partial_{\epsilon_0} Z_N(\cbeta)}{Z_N(\cbeta)}
\end{equation}
evaluated at complex temperatures. Fig.~2 displays contour plots of
$|\eta_0(N,\cbeta)|/N$ for 40, 120, and 300 particles.   The locations
of the zeros of $Z(\cbeta)$ (poles of $\eta_0(N,\cbeta)$) are indicated
by the white spots.   The separation of the condensed (dark) and the
normal (bright) phase is conspicuous. The zeros act like ''{\sl boundary
posts}'' between both phases. The boundary line between both phases gets
more and more pronounced as the number of particles increases and the
distance between neighboring zeros decreases. Fig.~2 virtually displays
how the phase transition approaches its thermodynamic limit. We have
determined the classification parameters for the phase transition by a
numerical analysis of the DOZ for up to 400 particles. The results are
given in Fig.~3. The parameter $\alpha$ is constant at about 1.25. The
small fluctuations are due to numerical errors in the determination of
the location of the zeros. This value of $\alpha$ indicates a third
order phase transition in the 3-dimensional harmonic trap. Results for
the 2-dimensional systems and other trap geometries, which will be
published elsewhere in detail,  indicate that the order of the phase
transition depends strongly on the trap geometry. The parameter $\gamma$
and the non-integer fraction of $\alpha$ are related to the critical
indices of the phase transition, e.g. $\gamma =0$ indicates equal
critical indices for approaching the critical temperature from the left
and from the right. Regarding the finite size effects $\tau_1$ is of
major importance. As can be seen in Fig.~3 (b) $\tau_1$ is approximately
proportional to $1/N$ that the systems of bosons in a 3 dimensional
harmonic trap approaches a true higher order phase transition linearly
with increasing particle number $N$.

It appears that the DOZ for Bose-Einstein condensates
is rather smooth. As an example for a little more complicated
situation we calculated the DOZ for small Ar clusters,
which have been extensively studied in the past
\cite{Berry:1984a,Berry:1984,Beck:1987,Jellinek:1986}. Their
thermodynamic behavior is governed by a {\sl hopping process} between
different isomers and {\sl melting} 
\cite{Labastie:1990,Wales:1989,Kunz:1994}. Many indicators of
{\sl phase transitions} in Ar clusters have been investigated, e.g.
the specific heat \cite{bgh96,hbsh96}, the rms bond length
fluctuation \cite{bor94a}, and the onset of an $1/f$-noise behavior of
the potential energy in time dependent molecular dynamics simulations
\cite{Nayak:1995}. However, for a good reason, all these 
attempts lack a definite classification of the transitions taking
place in these clusters.
Without going into the details of our numerical method which is based on
a determination of the interaction density of states by extensive Monte
Carlo simulations along with an optimized data analysis
\cite{Ferren:1989} we give here the results for Ar$_{6}$ and Ar$_{30}$.
Fig.~4 displays contour plots of the absolute value of the specific
heat $c_V(\cbeta)$ in the complex temperature plane. 
For Ar$_6$ the poles lie on a straight line at $T\simeq 15$~K and are
equally spaced with resulting classification parameters $\alpha=0$,
$\gamma=0$, and $\tau_1 \hbar = 0.05$ ps. From earlier works
\cite{fhb93} it is well known that at this temperature a hopping
transition between    the octahedral and the bicapped tetrahedral isomer
occurs. Our classification scheme now indicates that this isomer hopping
can be identified as a first order phase transition.  Ar$_{30}$ already
has a tremendous number of different isomers, and a much more
complicated form of the DOZ arises (see Fig.~4~(b)).  The DOZ cuts the
complex temperature plane into three regions with two transition lines
approaching the real axis. Comparing with the literature the region
below 31~K can be identified as the solid phase and the region above
45~K as a fluid phase. Because our MC simulations are performed at zero
pressure at this temperature also the evaporation of atoms from the
cluster starts which corresponds to the onset of the gas phase. The
phase between these two transition lines is commonly interpreted as the
melting, isomer hopping, or coexistance region.  The analysis of the
order of the phase transitions is quite difficult in this case and will
be investigated in a more systematic study.  Nevertheless the DOZ
displays in a distinct manner the phase separation for Ar$_{30}$ and can
be viewed as a unique fingerprint.

In conclusion we have found that the DOZ of the canonical partition
function can be used to classify phase transitions in finite systems.
The DOZ of a specific system acts like a unique fingerprint.  The
classification scheme given above is equivalent to that given by
Grossmann {\sl et al.} but extended to the region of finite particle
numbers. We have found that the zeros of the partition function act like
boundary posts between different phases in the complex temperature
plane. The finite size effects for the Bose-Einstein condensation are
reflected by an $1/N$ dependence of the parameter $\tau_1$ and only a
slight change of the parameter $\alpha$ which indicates the order of the
phase transition.  For Ar clusters the DOZ leads to enlightening
pictures of the complex process of melting or isomer hopping,
identifying in a distinct manner two critical temperatures supporting an
old assumption of Berry {\sl et al.}\cite{Berry:1984,Berry:1984a}.  This
classification scheme developed for the canonical ensemble
  should also
hold for other ensembles, i.e. different experimental conditions should
not influence the {\sl nature} of the systems although e.g. the shape of
the caloric curve may significantly differ. 

%
%
\begin{figure}
\centerline{\psfig{file=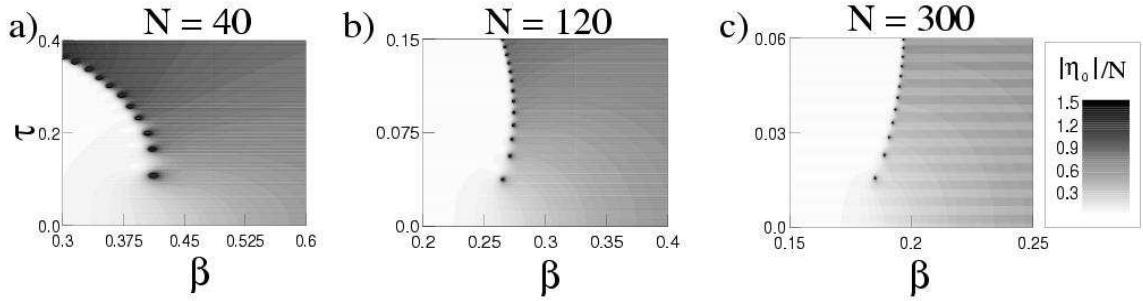,width=16cm}}
\caption{Contour plots the ground state
occupation number $\mid \eta_0 \mid /N$ in the complex temperature
plane for 40, 120, and 300 particles in a 3-dimensional isotropic trap.
The black spots indicate the locations of zeros of the partition
function.}
\label{figure2}
\end{figure}
\begin{figure}
\centerline{\psfig{file=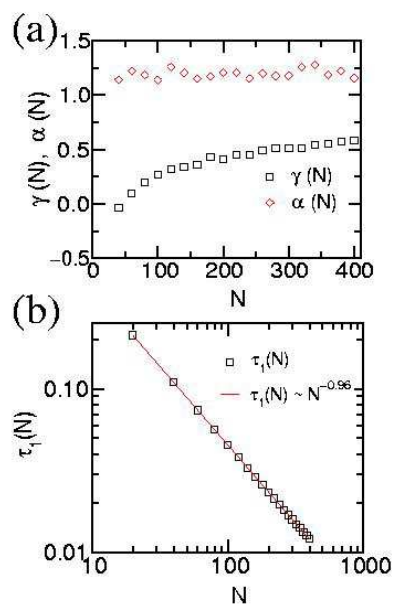,width=6cm}}
\caption{Plots of the classification parameters $\alpha$, $\gamma$,
and $\tau_1$ versus the number of particles for a 3-dimensional 
harmonic trap.}
\label{figure3}
\end{figure}
\begin{figure}
\centerline{\psfig{file=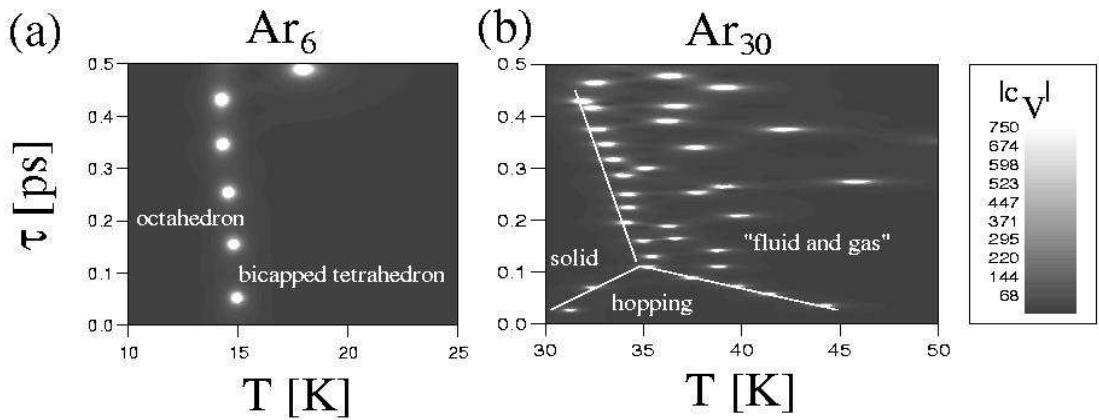,width=16cm}}
\caption{Contour plots of the specific heat $\mid c_V \mid$ for 
Ar$_6$ and Ar$_{30}$ clusters. }
\label{figure4}
\end{figure}
%
%


\begin{thebibliography}{27}
\bibitem{extension}
Phase transition with respect to other thermodynamic variables can 
be inspected by making these quantities complex an proceeding in the 
same way as for the temperature.
\end{thebibliography}
\end{document}